\documentclass[aps,prl,preprint,amsmath,showpacs]{revtex4-1}
\usepackage{bm}

\begin{document}

\title{Linear electro-optic effects due to high order spatial dispersion}

\author{F.~Castles}
\affiliation{Department of Materials, University of Oxford, Parks Road, Oxford OX1 3PH, United Kingdom}

\begin{abstract}

Two new types of electro-optic effect that are linear in the applied electric field strength are theoretically predicted to exist in transparent dielectric crystals due to high order spatial dispersion.  The first effect, which is quadratic in the wave vector of light, is possible in materials belonging to all noncentrosymmetric crystal classes.  The second, which is cubic in the wave vector, is possible in all crystals.  In the ${\bm O}(432)$ and $\bm{O_h}(m\bar 3m)$ crystal classes, for which the primary and secondary linear electro-optic effects and linear electrogyration are simultaneously absent, these effects lead, respectively, to qualitatively new behavior and constitute the dominant bulk electro-optic effect in the limit of small fields.  Thus, bulk linear electro-optic effects are predicted to exist in a wide range of materials---including many of considerable technological importance, such as silicon---where they were previously considered impossible.
\end{abstract}



\maketitle

It is well known that the optical properties of a material may be modified by the application of an electric field.  The discovery of, and fundamental research concerning, such phenomena has enabled the flow of light to be precisely manipulated, and ultimately led to many important technologies.  For example, electro-optic effects in polar liquids and solid crystals have been exploited in modulators for laser applications, and the electro-optic properties of liquid crystals are exploited in the majority of today's flat panel displays \cite{hecht,goldstein,handbookv8,fmi}.  Therefore, aside from any contribution to basic knowledge, the discovery of new electro-optic phenomena is of potential interest from an applied perspective.

Herein, electro-optic effects are considered in the conventional, but restricted, sense: as effects whereby an externally applied electric field which is static or varies slowly with respect to the frequency of light affects the refractive index of the material.  The refractive index is defined with respect to light propagation in the bulk of the homogeneous material, hence surface effects are not considered.  The analysis is further limited to nonmagnetic materials that are transparent within a given range of frequencies, which is the most important case in the theory of crystal optics \cite{landaubook}.

Within this domain, a general account of electro-optic effects is provided by expanding the material's dielectric tensor $\bm{\varepsilon}$, or inverse dielectric tensor $\bm{\varepsilon}^{-1}$, in powers of the applied electric field ${\bf E}$ \cite{nyebook,landaubook,agranovichbook}
\begin{equation} \label{eq:expansionepsilon}
\varepsilon_{jl}^{-1}(\omega,{\bf E},{\bf k}) = \eta_{jl}^{(0)}(\omega,{\bf k})+\eta_{jlm}^{(1)}(\omega,{\bf k})E_m + \eta_{jlmn}^{(2)}(\omega,{\bf k})E_mE_n + ... \, ,
\end{equation}
where $\omega$ is the frequency of the light and ${\bf k}$ is its wave vector in the material.  Terms in the expansion that are linear in $E$ are of particular interest in the respect that they formally dominate all effects of higher order in $E$ in the limit $E\rightarrow 0$.  The condition of small $E$, which is required in any case for expansion (\ref{eq:expansionepsilon}) to be useful, is physically relevant in many circumstances.  As a result, for example, electro-optic modulators that exploit effects linear in $E$ can operate at lower voltages than those that exploit effects quadratic in $E$, contributing to the replacement of quadratic `Kerr cells' by linear `Pockels cells' in most laser applications \cite{goldstein}.

A considerable limitation, however, is that electro-optic effects linear in $E$ are generally thought to be present only in certain materials with suitable macroscopic symmetries.  (For example, the lack of a linear electro-optic effect in silicon---which is a limitation in the context of silicon optoelectronics---was highlighted and redressed in \cite{jacobsen} by straining the silicon to break the $\bm{O_h}$ symmetry.)  Pockels investigated the primary and secondary linear electro-optic effects and argued that they are possible in materials belonging to 20 of the 32 crystal classes; they are forbidden in the 11 centrosymmetric crystal classes and the $\bm O$ crystal class \cite{pockels,nyebook,landaubook}.  However, Pockels's analysis ignores the effects of spatial dispersion \cite{landaubook,agranovichbook}, i.e., it ignores the dependence of $\bm\varepsilon$ or $\bm\varepsilon^{-1}$ on ${\bf k}$ at fixed $\omega$, which has already been indicated in expansion~(\ref{eq:expansionepsilon}).  It has previously been established, both theoretically and experimentally, that spatial dispersion leads to an electro-optic effect, known as electrogyration, that is linear in $E$ but is distinct from the primary or secondary linear electro-optic effects \cite{zheludevboth,zheludev1976both}.  Linear electrogyration is forbidden in only three of the 32 crystal classes: $\bm O$, $\bm{T_d}$, and $\bm{O_h}$ \cite{zheludevboth,zheludev1976both}.  This leaves two crystal classes, $\bm O$ and $\bm{O_h}$, in which the primary and secondary linear electro-optic effects and linear electrogyration are simultaneously absent.  Herein I address the question: are all bulk electro-optic effects linear in $E$ rigorously forbidden in transparent dielectrics belonging to the $\bm O$ and $\bm{O_h}$ crystal classes, or are new effects possible on account of higher order spatial dispersion?

For small ${\bf k}$ the components of the tensor ${\bm\eta}^{(0)}(\omega,{\bf k})$, which describe the optical properties of the material in the absence of an electric field, may be expanded in powers of ${\bf k}$ \cite{landaubook,agranovichbook}
\begin{equation} \label{eq:expansionnatural}
\eta_{jl}^{(0)}(\omega,{\bf k})=
\eta_{jl}^{(0,0)}(\omega)
+i\,\eta_{jln}^{(0,1)}(\omega)k_n
+\eta_{jlnp}^{(0,2)}(\omega)k_nk_p
+i\,\eta_{jlnpq}^{(0,3)}(\omega)k_nk_pk_q+ ...\, .
\end{equation}
The tensor $\bm\eta^{(0,0)}(\omega)$ describes the basic optical properties of the material in the absence of spatial dispersion (cubic, uniaxial, or biaxial) \cite{landaubook}, certain components of the tensor $\bm\eta^{(0,1)}(\omega)$ describe natural optical activity \cite{landaubook,agranovichbook}, the tensor $\bm\eta^{(0,2)}(\omega)$ describes, in particular, the intrinsic optical anisotropy of cubic crystals \cite{landaubook,agranovichbook}, and certain components of the tensor $\bm\eta^{(0,3)}(\omega)$ describe higher-order ($\propto k^3$) natural optical activity \cite{agranovich1,molchanovboth,agranovichbook}.

Similarly, the tensor ${\bm \eta}^{(1)}(\omega,{\bf k})$, which describes electro-optic effects linear in $E$, may be expanded \cite{agranovichbook}
\begin{equation} \label{eq:expansioneo}
\eta_{jlm}^{(1)}(\omega,{\bf k}) = 
\eta_{jlm}^{(1,0)}(\omega)
+i\,\eta_{jlmn}^{(1,1)}(\omega)k_n
+\eta_{jlmnp}^{(1,2)}(\omega)k_nk_p
+i\,\eta_{jlmnpq}^{(1,3)}(\omega)k_nk_pk_q
+...\, .
\end{equation}
The tensor $\bm\eta^{(1,0)}(\omega)$ describes the primary and secondary linear electro-optic effects \cite{pockels,nyebook}, and certain components of the tensor $\bm\eta^{(1,1)}(\omega)$ describe linear electrogyration \cite{agranovichbook}.

A general expansion of this type---i.e., of $\bm\varepsilon^{-1}$ in ${\bf k}$ and ${\bf E}$---was presented by Agranovich and Ginzburg in \cite{agranovichbook}, p. 194; the higher-order electro-optic terms included in expansion~(\ref{eq:expansioneo}) that are $\propto k^2$ and $\propto k^3$ are, of course, implicit in Agranovich and Ginzburg's theory, but were not included explicitly or discussed in \cite{agranovichbook}.  Nor, to my knowledge, have they been included or discussed elsewhere in the literature.  One may naively assume that such terms are relatively unimportant in the respect that they will be largely masked by lower order, more dominant, effects.  Indeed, in most materials this will be the case.  However, what I argue is that, in the $\bm O$ and $\bm{O_h}$ crystal classes, these terms lead to qualitatively and substantially new behavior: the appearance of electro-optic effects linear in $E$ where otherwise there are none.  In the limit $E\rightarrow 0$, these will always formally constitute the dominant electro-optic effect.

The condition of no absorption imposes the relation $\varepsilon_{jl}^{-1}(\omega,{\bf E},{\bf k})=[\varepsilon_{lj}^{-1}(\omega,{\bf E},{\bf k})]^*$, where $^*$ denotes complex conjugation, and the generalized principle of the symmetry of the kinetic coefficients imposes the relation $\varepsilon_{jl}^{-1}(\omega,{\bf E},{\bf k})=\varepsilon_{lj}^{-1}(\omega,{\bf E},-{\bf k})$ \cite{landaubook,agranovichbook}.  It follows that all tensor components on the right hand side of expansions~(\ref{eq:expansionnatural}) and (\ref{eq:expansioneo}) are real (factors of $i$ were inserted to ensure certain terms are real rather than pure imaginary) and the relations between the components of these tensors may be denoted
\begin{equation} \nonumber
\eta_{(jl)}^{(0,0)}(\omega), 
\quad \eta_{[jl]n}^{(0,1)}(\omega), 
\quad \eta_{(jl)(np)}^{(0,2)}(\omega), 
\quad \eta_{[jl](npq)}^{(0,3)}(\omega),
\end{equation}
\begin{equation}
\eta_{(jl)m}^{(1,0)}(\omega), 
\quad \eta_{[jl]mn}^{(1,1)}(\omega), 
\quad \eta_{(jl)m(np)}^{(1,2)}(\omega), 
\quad\textrm{and}\quad \eta_{[jl]m(npq)}^{(1,3)}(\omega),
\end{equation}
where rounded brackets in the subscripts denote symmetry with respect to permutation of the contained indices, e.g., $\eta_{(jl)m}^{(1,0)}(\omega)$ implies $\eta_{jlm}^{(1,0)}(\omega)=\eta_{ljm}^{(1,0)}(\omega)$, and square brackets denote antisymmetry, e.g., $\eta_{[jl]mn}^{(1,1)}(\omega)$ implies $\eta_{jlmn}^{(1,1)}(\omega)=-\eta_{ljmn}^{(1,1)}(\omega)$.

The macroscopic symmetries of a particular material may impose additional relations between the components of a given tensor which further restrict the number of independent components it contains.  In some instances the symmetry relations are so restrictive that all components of a given tensor must be simultaneously zero, in which case we may say that the physical effect associated with the tensor is forbidden by symmetry.  This constitutes a powerful technique by which one may ascertain if a given physical effect is possible in a given material, based only on the macroscopic symmetry of the material and without regard to its microscopic composition \cite{nyebook}.

In the $\bm O$ crystal class all the components of $\bm\eta^{(1,0)}(\omega)$ are simultaneously zero, thus no primary or secondary linear electro-optic effect is possible \cite{pockels,nyebook,landaubook}.  $\bm\eta^{(1,1)}(\omega)$ contains one independent component which, however, does not enter the relevant wave equation [Eq.~(\ref{eq:wavemain}) below] and does not affect the refractive index.  Thus, it does not constitute an electro-optic effect in the sense considered herein (though it may be observed via reflection \cite{novikovboth} and may, in that regard, be considered a surface effect).  This is consistent with the known lack \cite{zheludevboth,zheludev1976both} of electrogyration in $\bm O$.  The tensor $\bm\eta^{(1,2)}(\omega)$ contains three independent components (herein such results are arrived at using \textit{Ten$\chi$ar} software \cite{*[{}] [{.  Software currently available for free download via http://it.iucr.org/Da/resource1/ .}] tenxar}), indicating that an electro-optic effect that is linear in $E$ is possible in $\bm O$.

In general, the components of $\bm\eta^{(1,2)}(\omega)$ may be grouped with the intrinsic $\propto k^2$ terms
\begin{equation}
\varepsilon_{jl}^{-1}(\omega,{\bf E},{\bf k}) = ...+ \left[\eta_{(jl)(np)}^{(0,2)}(\omega)+\eta^{(1,2)}_{(jl)m(np)}(\omega)E_m \right]k_nk_p +... \, , 
\end{equation}
and their effect may be thus interpreted as an electrically-induced modulation in the $\propto k^2$ optical properties of the crystal (in particular, as an electrically-induced modulation in the $\propto k^2$ optical anisotropy of cubic crystals).  The tensor $\bm\eta^{(1,2)}(\omega)$ contains nonzero components in the 21 noncentrosymmetric crystal classes, while all components are simultaneously zero in the 11 centrosymmetric ones.  The $\bm O$ crystal class is of singular interest because it is the only noncentrosymmetric class for which the primary and secondary linear electro-optic effects are forbidden.  Thus, because electrogyration is also absent in $\bm O$, the $\propto Ek^2$ effect, governed by the tensor $\bm\eta^{(1,2)}(\omega)$, becomes the lowest-order electro-optic effect.

The specific nature of the $\propto Ek^2$ electro-optic effect may be investigated in more detail by looking at the effect of $\bm\eta^{(1,2)}$ in the wave equation.  It is convenient to consider a Cartesian coordinate system which has one axis, $\hat{\bf x}_3$ say, along ${\bf k}$, for which the wave equation may be written \cite{landaubook}
\begin{equation} \label{eq:wavemain}
\left[\frac{1}{n^2}\,\delta_{\alpha\beta}-\epsilon^{-1}_{\alpha\beta}(\omega,{\bf E},{\bf k})\right]D_\beta=0,
\end{equation}
where $n=ck/\omega$ is the refractive index ($c$ is the speed of light in vacuum and $k=|{\bf k}|$), $\delta_{\alpha\beta}$ is Kronecker's delta, $D_\beta$ are the components of the electric induction, and the Greek suffixes take the values 1 or 2 corresponding to the axes $\hat{\bf x}_1$ and $\hat{\bf x}_2$.  Note that $D$ refers here to the light wave whereas ${\bf E}$ refers to the external electric field, and the analysis is restricted to nonlongitudinal waves.  To create a simple but nontrivial example we may consider the scenario where ${\bf E}$ is parallel to ${\bf k}$ and both are along a 2-fold rotation axis of the $\bm O$ structure.  Let $\hat{\bf x}_1$ and $\hat{\bf x}_2$ lie along 4-fold and 2-fold rotation axes respectively.  For clarity, we may artificially let $\bm\eta^{(0,1)}(\omega)=\bm 0$ and thus ignore the effects of natural optical activity.  In this case the wave equation may be written \cite{suppPRL}
\begin{equation}\label{eq:waveo}
\left(
\begin{array}{cc}
\frac{1}{n^2}-\eta'^{(0,0)}(\omega)-\eta'^{(0,2)}_{A}(\omega)k & -\eta'^{(1,2)}(\omega)k^2E \\
-\eta'^{(1,2)}(\omega)k^2E & \frac{1}{n^2}-\eta'^{(0,0)}(\omega)-\eta'^{(0,2)}_{B}(\omega)k
\end{array}
\right)
\left(
\begin{array}{c}
D_1 \\
D_2 \\
\end{array}
\right)
=
\left(
\begin{array}{c}
0 \\
0 \\
\end{array}
\right),
\end{equation}
where $\eta'^{(0,0)}(\omega)$ is the only independent component of the tensor $\bm\eta^{(0,0)}(\omega)$, $\eta'^{(0,2)}_{A}(\omega)$ and $\eta'^{(0,2)}_{B}(\omega)$ are the two independent components of $\bm\eta^{(0,2)}(\omega)$ that enter the wave equation in this configuration, and $\eta'^{(1,2)}(\omega)$ is the one independent component of $\bm\eta^{(1,2)}(\omega)$ that enters the wave equation in this configuration.  The one independent component of $\bm\eta^{(1,1)}(\omega)$ has not entered the wave equation, consistent with the discussion above.  While it is clear that $E$ can affect the refractive indices via solution of the characteristic equation associated with Eq.~(\ref{eq:waveo}), it is more pertinent to consider the effect of $E$ on the polarization of the eigenwaves.  In the absence of the electric field, Eq.~(\ref{eq:waveo}) reduces to the scenario considered in \cite{agranovichbook} where $\eta'^{(0,2)}_A(\omega)$ and $\eta'^{(0,2)}_B(\omega)$ introduce anisotropy in the cubic structure, assuming they are unequal.  In this field-free case the two eigenwaves are linearly polarized: one along $\hat{\bf x}_1$ and the other along $\hat{\bf x}_2$.  The effect of the electric field may be elucidated by noting that in the limit of small $E$, Eq.~(\ref{eq:waveo}) may be written
\begin{equation}
\left(
\begin{array}{cc}
\frac{1}{n^2}-\eta'^{(0,0)}(\omega)-\eta'^{(0,2)}_A(\omega)k & 0 \\
0 & \frac{1}{n^2}-\eta'^{(0,0)}(\omega)-\eta'^{(0,2)}_B(\omega)k
\end{array}
\right)
\left(
\begin{array}{c}
D'_1 \\
D'_2 \\
\end{array}
\right)
=
\left(
\begin{array}{c}
0 \\
0 \\
\end{array}
\right),
\end{equation}
where
\begin{equation}
\left(
\begin{array}{c}
D'_1 \\
D'_2 \\
\end{array}
\right)
=
\left(
\begin{array}{cc}
\cos\theta & \sin\theta \\
-\sin\theta & \cos\theta 
\end{array}
\right)
\left(
\begin{array}{c}
D_1 \\
D_2 \\
\end{array}
\right)
,\quad
\theta=\frac{\eta'^{(1,2)}(\omega)kE}{\eta'^{(0,2)}_A(\omega)-\eta'^{(0,2)}_B(\omega)},
\end{equation}
and we have assumed that $\eta'^{(0,2)}_A(\omega)\neq\eta'^{(0,2)}_B(\omega)$.  We see that the equation takes the same form as the field-free case except that ${\bf D}$ is rotated.  That is, in the limit of small $E$, upon application of an electric field the refractive index is approximately unchanged and the eigenwaves remain linearly polarized; the pertinent effect is that the orientation of the linear polarization of the eigenwaves is rotated about the $\hat{\bf x}_3$ axis.  The angle of rotation $\theta$ is proportional to $Ek$.

In the $\bm{O_h}$ crystal class all the components of $\bm\eta^{(1,0)}(\omega)$ are all simultaneously zero \cite{pockels,nyebook,landaubook}, and $\bm\eta^{(1,1)}(\omega)$ contains one independent component which, as before, does not enter the wave equation or affect the refractive index.  The components of $\bm\eta^{(1,2)}(\omega)$ are all simultaneously zero.  $\bm\eta^{(1,3)}(\omega)$ contains three independent components; since $\bm\eta^{(1,3)}(\omega)$, like $\bm\eta^{(1,1)}(\omega)$, pertains to the antisymmetric part of $\bm\varepsilon^{-1}(\omega,{\bf E},{\bf k})$, care must be taken in determining whether it can affect the wave equation.  To ascertain this, the general arguments given in Ref.~\cite{landaubook}, p. 365, Ref.~\cite{agranovichbook}, p. 127, and Ref.~\cite{molchanovboth} are here extended:  Consider the pseudovector ${\bf f}(\omega,{\bf E},{\bf k})$ that is defined from the antisymmetric part of $\bm\varepsilon^{-1}(\omega,{\bf E},{\bf k})$ according to
\begin{equation}
i\,e_{[jlr]}f_r(\omega,{\bf E},{\bf k}) = \frac{1}{2}\left[\varepsilon^{-1}_{jl}(\omega,{\bf E},{\bf k})-\varepsilon^{-1}_{lj}(\omega,{\bf E},{\bf k})\right],
\end{equation}
where ${\bf e}$ is the completely antisymmetric unit pseudotensor of rank three.  In the general case this gives
\begin{equation} \label{eq:f}
f_r(\omega,{\bf E},{\bf k}) = \left[\Lambda_{rn}^{(0,1)}(\omega)+\Lambda_{rnm}^{(1,1)}(\omega)E_m\right]k_n + \left[\Lambda_{r(npq)}^{(0,3)}(\omega)+\Lambda_{r(npq)m}^{(1,3)}(\omega)E_m\right]k_nk_pk_q
+\,...\, ,
\end{equation}
where the $\bm\Lambda$s are pseudotensors dual to the $\bm\eta$-tensors
\begin{equation}
\begin{array}{ccc}
e_{[jlr]}\Lambda_{rn}^{(0,1)}(\omega)=\eta^{(0,1)}_{[jl]n}(\omega), & &
e_{[jlr]}\Lambda_{rnm}^{(1,1)}(\omega)=\eta^{(1,1)}_{[jl]mn}(\omega), \\
e_{[jlr]}\Lambda_{r(npq)}^{(0,3)}(\omega)=\eta^{(0,3)}_{[jl](npq)}(\omega), & \quad\textrm{and}\quad\quad &
e_{[jlr]}\Lambda_{r(npq)m}^{(1,3)}(\omega)=\eta^{(1,3)}_{[jl]m(npq)}(\omega).
\end{array}
\end{equation}
Only the projection of ${\bf f}(\omega,{\bf E},{\bf k})$ along ${\bf k}$, i.e., the scalar product $k_jf_j(\omega,{\bf E},{\bf k})$, enters the wave equation.  Thus, only the `symmetric parts' of the $\bm\Lambda$-pseudotensors which may be denoted
\begin{equation}
\Lambda_{(rn)}^{(0,1)s}(\omega),\quad
\Lambda_{(rn)m}^{(1,1)s}(\omega),\quad
\Lambda_{(rnpq)}^{(0,3)s}(\omega),\quad
\textrm{and}\quad
\Lambda_{(rnpq)m}^{(1,3)s}(\omega),
\end{equation}
enter the wave equation.  The question of whether any of the nonzero components of the tensor $\bm\eta^{(1,3)}(\omega)$ can enter the wave equation is now equivalent to asking whether the pseudotensor $\bm\Lambda^{(1,3)s}(\omega)$ contains any nonzero components.  In the $\bm{O_h}$ crystal class the answer is that the pseudotensor $\bm\Lambda^{(1,3)s}(\omega)$ contains one independent component.  Thus, of the three independent components of $\bm\eta^{(1,3)}(\omega)$ in $\bm{O_h}$, one can enter the wave equation.  On account of this component, an electro-optic effect that is linear in $E$ is indeed possible in the $\bm{O_h}$ crystal class.

In general, the tensor $\bm\Lambda^{(1,3)s}(\omega)$ describes the creation or modulation of higher-order natural optical activity of the type reported in \cite{molchanovboth}.  The effect may be interpreted as a $\propto Ek^3$ analogue of electrogyration (electrogyration being $\propto Ek$).  The components of $\bm\eta^{(1,3)}(\omega)$ that do not enter the wave equation do not constitute an electro-optic effect in the sense considered herein but describe a `weak' effect which is a higher-order analogue of `weak gyrotropy' (Ref.~\cite{agranovichbook}, p. 128, and Ref.~\cite{fedorov2both}) and the effect reported in Ref.~\cite{novikovboth}.  Both $\bm\eta^{(1,3)}(\omega)$ and $\bm\Lambda^{(1,3)s}(\omega)$ contain nonzero components in all crystal classes.  The $\bm{O_h}$ crystal class is of singular interest with respect to the $\propto Ek^3$ effect because it is the only class that does not admit effects $\propto E$ that are of lower order in $k$.  Thus the $\propto Ek^3$ effect, governed by the tensor $\bm\eta^{(1,3)}(\omega)$, becomes the lowest-order electro-optic effect in $\bm{O_h}$.

The wave equation in an $\bm{O_h}$ material is now considered for a specific example configuration.  Let ${\bf E}$ be directed along a 4-fold symmetry axis of the cubic structure and let ${\bf k}$ be perpendicular to ${\bf E}$ and at $\pi/8$ to a 4-fold axis.  Choosing $\hat{\bf x}_1$ along ${\bf E}$, the wave equation, Eq.~(\ref{eq:wavemain}), may be written \cite{suppPRL}
\begin{equation}
\left(
\begin{array}{cc}
\frac{1}{n^2}-\eta'^{(0,0)}(\omega)-\eta'^{(0,2)}_{A}(\omega)k^2 
& \quad -i\,\eta'^{(1,3)}(\omega)Ek^3\quad \\
i\,\eta'^{(1,3)}(\omega)Ek^3
& \quad\frac{1}{n^2}-\eta'^{(0,0)}(\omega)-\eta'^{(0,2)}_{B}(\omega)k^2
\end{array}
\right)
\left(
\begin{array}{c}
D_1 \\
D_2 \\
\end{array}
\right)
=
\left(
\begin{array}{c}
0 \\
0 \\
\end{array}
\right).
\end{equation}
Note that this explicitly confirms that a component of $\bm\eta^{(1,3)}$ can enter the wave equation.  By comparison with standard analyses \cite{landaubook,agranovichbook}, the pure imaginary nature of the $\propto Ek^3$ term and its position in the off-diagonal elements of the matrix associate its effect with optical activity.  For $E=0$, the eigenwaves are linearly polarized and nonzero $E$ introduces ellipticity.

It may be noted in passing that effects $\propto Ek^n$ for $n\geq4$ are also formally indicated by expansion (\ref{eq:expansioneo}), but they are relatively unimportant in the respect that they will always be accompanied, and largely masked, by lower order, more dominant, effects.

The line of reasoning developed herein was stimulated by the recent observation of an electro-optic response linear in $E$ in a liquid crystal blue phase \cite{castlesnatmat2} whose equilibrium structure is known \cite{kitzerow3} to belong to the $\bm O$ crystal class.  Given the known lack of primary and secondary linear electro-optic effects and electrogyration in $\bm O$, the possibility of new electro-optic effects induced by spatial dispersion was initially considered among other working hypotheses.  In fact, further experiments revealed that the observations were consistent with strain in the sample reducing the symmetry of the structure to a crystal class that permits a primary linear electro-optic effect \cite{castlesnatmat2}.  Nevertheless, the idea of new electro-optic effects remained theoretically viable and the arguments presented herein stand as general theoretical predictions which the experiments reported in \cite{castlesnatmat2} were neither sensitive nor rigorous enough to detect.  The cubic blue phases may yet be a promising material in which to search for experimental evidence of the $\propto Ek^2$ effect, if suitably sensitive and rigorous experimental methods were employed.

A number of technologically important materials such as crystalline silicon belong to the $\bm{O_h}$ crystal class.  Thus, the $\propto Ek^3$ effect may have significance, for example, in the field of silicon optoelectronics.  While the effect will always formally constitute the dominant electro-optic effect for $E\rightarrow 0$, according to expansions (\ref{eq:expansionepsilon}) and (\ref{eq:expansioneo}), it must be pointed out that, loosely speaking, the magnitude of the effect will be small (which would explain why it has not previously been observed).

It is natural to refer to all the above electro-optic effects that are linear in $E$ as linear electro-optic effects, of which the effects investigated by Pockels \cite{pockels} are an archetypal subset.  Accordingly, we have arrived at the conclusion that linear electro-optic effects of one kind or another are possible in all transparent crystalline dielectrics \cite{*[{A similar conclusion was drawn in }][{.  The effect considered there, which is $\propto Ek$, does not entail any change in the refractive index and is not an electro-optic effect in the conventional sense considered herein.}] bali}.  Since the theory concerns only the macroscopic symmetries of the material, without regard to the microscopic origin of the effects, the `optical' properties of `crystalline' materials may be interpreted in a broad sense.  That is, the analysis applies equally to all spatially periodic media---atomic and molecular crystals, liquid crystals (cubic blue phases, for example), and dielectric crystals (photonic crystals or all-dielectric metamaterials)---and all wavelengths of electromagnetic radiation, provided the wavelength is sufficiently large with respect to the periodicity of the lattice that a macroscopic $\bm\varepsilon$ or $\bm\varepsilon^{-1}$ may be appropriately defined.

\medskip
This work was funded by the Engineering and Physical Sciences Research Council UK (grant EP/I034548/1).

\end{document}